\documentclass[sigconf]{acmart}
\usepackage{blindtext}
\usepackage{makecell} 
\usepackage{graphicx}

\usepackage{subcaption}

\usepackage{caption}

\usepackage[compact]{titlesec} 

\AtBeginDocument{%
  }

\setcopyright{rightsretained}

\setcctype{by}

\begin{document}

\title[Towards Real-Time Detection of Inaccurate PPG Heart Rate]{Towards a Real-Time Warning System for Detecting Inaccuracies in Photoplethysmography-Based Heart Rate Measurements in Wearable Devices}


\settopmatter{authorsperrow=3}

\author{Rania Islambouli}
\orcid{0000-0003-1689-8938}
\affiliation{%
  \institution{Ludwig Boltzmann Institute for Digital Health and Prevention}
  \city{Salzburg}
  \country{Austria}}
\email{rania.islambouli@dhp.lbg.ac.at}

\author{Marlene Brunner}
\affiliation{%
  \institution{The Salzburg University of Applied Sciences}
  \city{Salzburg}
  \country{Austria}}
\email{ marlene.brunner@fh-salzburg.ac.at}

\author{Devender Kumar}
\orcid{0000-0002-6971-2829}
\affiliation{%
  \institution{University of Southern Denmark
}
  \city{Odense}
  \country{Denmark}}
\email{deku@mmmi.sdu.dk}

\author{Mahdi Sareban}
\orcid{0000-0002-8146-0505}
\affiliation{%
  \institution{Institute of Sports Medicine, Prevention and Rehabilitation
Paracelsus Medical University}
  \city{Salzburg}
  \country{Austria}}
\email{m.sareban@salk.at}

\author{Gunnar Treff}
\orcid{0000-0001-5388-4282}
\affiliation{%
  \institution{Paracelsus Medical University}
  \city{Salzburg}
  \country{Austria}}
\email{gunnar.treff@pmu.ac.at}

\author{Michael Neudorfer}
\affiliation{%
  \institution{Institute of Sports Medicine, Prevention and Rehabilitation
Paracelsus Medical University}
  \city{Salzburg}
  \country{Austria}}
\email{m.neudorfer@salk.at}

\author{Josef Niebauer}
\affiliation{%
  \institution{Ludwig Boltzmann Institute for Digital Health and Prevention}
  \city{Salzburg}
  \country{Austria}}
\email{
josef.niebauer@lbg.ac.at}

\author{Arne Bathke}
\affiliation{%
  \institution{Paris Lodron University of Salzburg}
  \city{Salzburg}
  \country{Austria}}
\email{ arne.bathke@plus.ac.at}

\author{Jan David Smeddinck}
\orcid{0000-0003-0562-8473}
\affiliation{%
  \institution{Ludwig Boltzmann Institute for Digital Health and Prevention}
  \city{Salzburg}
  \country{Austria}}
\email{ jan.smeddinck@dhp.lbg.ac.at}
\renewcommand{\shortauthors}{Generel Chair et al.}

\begin{abstract}
Wearable devices with photoplethysmography (PPG) sensors are widely used to monitor heart rate (HR), yet often suffer from accuracy issues. However, users typically do not receive an indication of potential measurement errors. We present a real-time warning system that detects and communicates inaccuracies in PPG-derived HR, aiming to enhance transparency and trust. Using data from Polar and Garmin devices, we trained a deep learning model to classify HR accuracy using only the derived HR signal. The system detected over 80\% of inaccurate readings. By providing interpretable, real-time feedback directly to users, our work contributes to HCI by promoting user awareness, informed decision-making, and trust in wearable health technology.
\end{abstract}




\maketitle

\section{Introduction}
\label{section:intro}
Reliable and accurate heart rate (HR) monitoring is essential for assessing cardiovascular health~\cite{mejia2020pulse}, guiding physical training~\cite{visseren20222021, zhang2014troika}, and detecting potential health issues~\cite{hayano2022enhanced, achten2003heart}.  Currently, electrocardiograms (ECGs) are the gold standard for HR measurement, providing valid and reliable data by detecting the electrical activity of the heart~\cite{schafer2013accurate}. However, the growing demand for continuous, non-invasive HR monitoring in everyday life~\cite{nittas2019electronic, hao2024advancing} has led to the rise of wearable devices equipped with photoplethysmography (PPG) sensors. Unlike ECGs, which require electrodes, PPG sensors estimate HR by detecting blood flow changes using light absorption~\cite{daimiwal2014respiratory}. This technology is now standard in consumer wearables such as smartwatches and fitness trackers, offering the convenience of continuous monitoring without complex or costly equipment~\cite{weiler2017wearable}. Despite this, PPG sensors are prone to inaccuracies~\cite{valopti1}, influenced by factors like motion artifacts, skin tone variations, and ambient light~\cite{bent2020investigating}. As a result, HR readings can differ significantly from ECG-based measurements~\cite{weiler2017wearable}, posing challenges for research and healthcare applications where precision is critical~\cite{biswas2019heart}. Inaccurate readings can also cause user anxiety~\cite{valins1966cognitive} and undermine trust in digital health tools.

To address this problem, researchers have focused on minimizing HR measurement errors and enhancing the accuracy of HR estimation. Previous approaches have employed various signal processing algorithms to reconstruct raw PPG sensor signals and eliminate inaccuracies~\cite{pankaj2022review}. Other methods have attempted to mitigate the impact of motion artifacts by integrating multiple PPG signals or combining accelerometer and PPG data~\cite{nabavi2020robust,seok2021motion}. However, these methods often require substantial computational power and introduce latency, making real-time processing difficult~\cite{pankaj2022review}. Moreover, they assume that the sensor Application Programming Interface (API) provides the corresponding waveforms from which HR measurements are derived, which is not always the case. Also, these techniques do not completely eliminate inaccuracies, underscoring the need for a system that can detect and report inaccuracies in real-time. If reliable decision-support algorithms are to be developed and used for patient monitoring, it is critical to distinguish between accurate and inaccurate data. It is also essential to be able to inform users in (near) real-time about inaccuracies.

In this work, we explore the viability of detecting periods of notably reduced accuracy of PPG-derived HR based only on HR signal derived from a given wearable at runtime, and we present a proof-of-concept warning system designed to generate information about the extent of inaccuracy that can be used to inform users. Our solution is independent of data collection hardware and HR estimation algorithms and does not require the analysis of the original waveforms. The system provides immediate feedback to users, allowing them to assess the accuracy of the measurements, take corrective actions, or seek more accurate measurements. Our contributions are threefold. First, we explore the feasibility of training a machine learning model with data from wearable PPG devices and ECG references to predict periods of reduced HR accuracy (\textbf{C1}). Second, we propose a prototypical real-time warning system that detects inaccuracies using a deep learning model, independent of hardware or waveform access (\textbf{C2}). Finally, we evaluate the system on data from popular Polar and Garmin devices and demonstrate that it can detect over 80\% of inaccurate measurements (\textbf{C3}).

\begin{figure*}[t]
    \centering
    \includegraphics[width=0.9\textwidth]{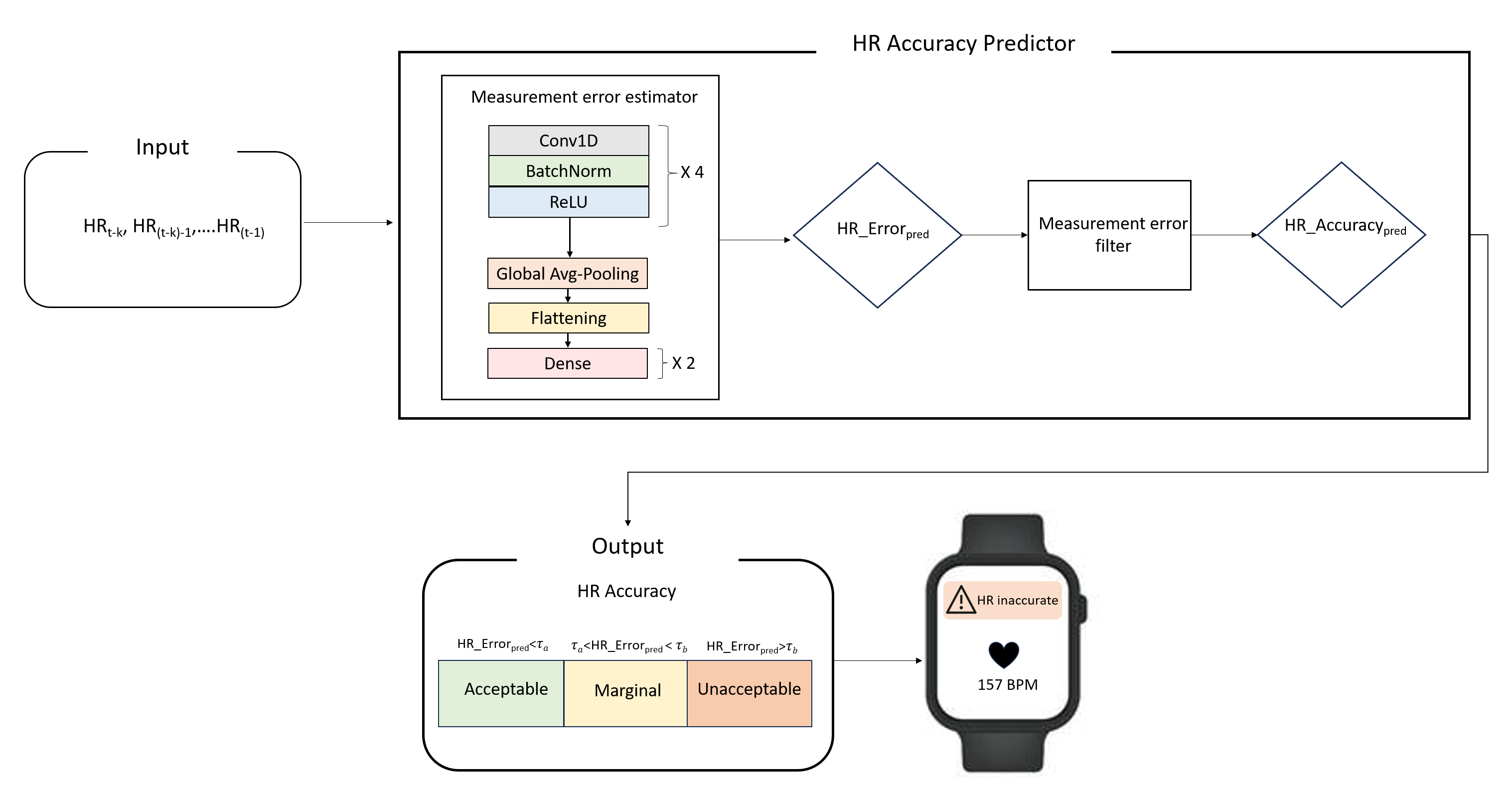}
    \caption{Real-Time Warning System for HR Accuracy: From PPG-Based HR Input Through CNN-Based Error Estimation to Threshold-Driven Accuracy Classification and User Feedbac\(t\).}
        \Description{A diagram showing a two-stage system for detecting heart rate inaccuracies from wearable devices. The first stage, the Measurement Error Estimator, takes recent PPG-derived heart rate values and predicts the measurement error using a convolutional neural network. The second stage, the Measurement Error Filter, compares the predicted error to two thresholds to classify the measurement as acceptable (green), marginal (yellow), or unacceptable (orange). The classification result is shown alongside the current heart rate reading to the user.}
    \label{fig:warning_system}
\end{figure*}

\section{Real-time warning system}
\label{sec:warning_system}
The proposed warning system is designed to provide real-time feedback on the accuracy of HR measurements obtained from PPG sensors. The system architecture, presented in Figure 1, is divided into two components: the \emph{Measurement Error Estimator} and the \emph{Measurement Error Filter}. The system operates by first predicting the potential error in HR measurements and then classifying the accuracy of those measurements based on the predicted error. For the final output in this first prototypical version of the system, we chose a color-coded indication of the accuracy level of the measurement based on a three-tier traffic light system, which is displayed alongside the HR reading on the wearable device.

\subsection{Measurement error estimator }
The measurement error estimator is responsible for predicting the HR measurement error at a given time $t$. It takes as input the sequence of the last $k$ HR measurements captured by the PPG sensor at a sampling rate of 1 
Hz. The estimator outputs a predicted HR measurement error, with the true error being the absolute difference between the HR recorded by the PPG sensor and the actual HR measured by a more accurate reference, in our application, a gold standard ECG sensor. This level of abstraction is interesting to consider as it can be independent of the underlying sensor and pre-processing, yet temporal patterns may still transport detectable signals of unreliable periods. Resulting from iterative design and testing with common machine learning methods and model parameterizations with the goal of establishing a proof-of-concept rather than a provably optimal system, the architecture of the measurement error estimator is depicted in Figure~\ref{fig:warning_system}. It consists of $4$ \emph{Convolution Neural Network (CNN) layers}. By applying convolutions across the temporal dimension of the sensor-derived HR data, the model can capture patterns and relationships within the time series data. Following the CNN layers, a \emph{global average pooling layer} is applied to condense the extracted features. This pooling process simplifies the data by averaging the values in the feature map, effectively reducing the complexity of the data while retaining its most critical characteristics. The output from this pooling layer is then passed through a \emph{flattening layer}, which transforms the multi-dimensional data into a single one-dimensional vector. This flattened vector is subsequently processed by two \emph{dense layers}, which together produce the final prediction of the HR measurement error. This simple architecture allows the model to be deployed on wearable devices with limited computational resources while having the opportunity to capture relevant temporal dynamics or patterns in the sensor data.

\subsection{Measurement error filter}
Once the measurement error is predicted by the estimator, it is passed to the \emph{measurement error filter} (MEF). This filter classifies the accuracy of the HR measurement into one of three categories: acceptable, marginal, or unacceptable. These categories were defined based on reasonable medical and biophysiological considerations, where small deviations (e.g., <10 bpm) are generally tolerable, moderate deviations may impact interpretation but are not necessarily critical, and large deviations (>30 bpm) are typically considered unreliable in both clinical and everyday contexts. As shown in Figure ~\ref{fig:warning_system}, the classification is determined by comparing the predicted measurement error against two predefined error thresholds $\tau_a$ and $\tau_b $, which delineate the boundaries for each accuracy level. Depending on the system use case and intended purpose of application, the levels can be set on the basis of critical system performance parameters (cf. Section 4.1), or on the basis of arguments around physiological relevance.The output is a color-coded accuracy label: 
\textbf{Acceptable (Green)} – HR is within a safe and accurate range; 
\textbf{Marginal (Yellow)} – HR is slightly outside the preferred accuracy range and requires caution; 
\textbf{Unacceptable (Orange)} – HR is significantly inaccurate and should not be trusted.

\section{System Implementation }
\label{sec:implemenation}
\subsection{Study Design and Dataset description}
The dataset utilized in this work for training the HR measurement estimator model originates from the the ValOpti Study~\cite{ValOpti, valopti1} which was designed to assess the validity and reliability of consumer-grade optical HR sensors in evaluating physical activity volume. It involved the collection of continuous HR data from participants using both a medical-grade ECG Holter monitor (Amedtec ECGpro, Aue, Germany) and four consumer-grade HR monitors: Garmin Venu 2S, Polar Verity Sense, Polar Vantage M2, and Scosche Rhythm24. The data was collected over
24 hours from 32 subjects at a frequency of 1Hz, with an average age of 59.1 ± 10.1, of whom 66\% were male. This study focused on establishing feasibility using the most accurate device, which was the Polar Verity Sense (arguably providing the most challenging use case for the system presented in this paper). Additionally, the study explored the transferability of the concept to other devices by incorporating data from the Garmin Venu2s. This device was chosen, because we wanted to include a non-Polar device as well and Scosche recordings were incomplete. 

\subsection{Dataset preprocessing}

The HR data recorded at 1Hz from the ECG Holter and the Polar and Garmin devices were synchronized based on their Universal Time Coordinated (UTC) timestamps. Following that, an additional step was performed using the Root Mean Square Error (RMSE) method to address any initial lag or discrepancies caused by differences in the clock frequencies of the recording devices, thereby minimizing time lags and ensuring precise synchronization. After synchronization, the ground truth HR data from the ECG Holter monitor were manually reviewed by a cardiologist to identify any outliers or implausible values. These were removed from the dataset and the associated timestamps were dropped. Any missing measurements in either the Polar or Garmin devices or the ECG Holter data were also removed. To prepare the HR data for input into the supervised learning model, data from the Polar and Garmin devices (referred to as PPG HR) were transformed into a rolling window format. Through exploratory testing, we considered window sizes ranging from 5 to 30 seconds to determine a configuration that could capture physiologically relevant patterns without compromising real-time applicability. Ultimately, we selected a 10-second window with a 9-second overlap and a 1-second slide. This configuration was chosen to establish the feasibility of the proof of concept, enabling the model to capture meaningful temporal patterns within short time frames rather than aiming for an optimal setup. 80\% of the data was allocated to the training set and 20\% to the test set. The data from subjects 1 to 25 were used for training, while data from subjects 26 to 32 were reserved for testing. This approach resulted in a total of 2,120,672 training windows and 598,596 test windows. 

\subsection{Training methodology}
The measurement error estimator was trained using the Adam optimizer with a learning rate of 0.001. A custom loss function was used, based on the absolute difference between the actual difference (diff\_true) between the ECG and wearable HR measurements and difference predicted by the model (diff\_pred). To find the best model configuration, we used GridSearch to explore and select the optimal hyperparameters. The model was trained over 200 epochs, with early stopping applied to avoid overfitting by halting training when model performance stopped improving. The CNN architecture used a kernel size of 3. The first two CNN layers had 8 neurons each, and the third and fourth layers had 16 neurons each. 

\subsection{Testing Methodology}
To evaluate our model, we set a threshold for the measurement error filter to classify the accuracy of HR measurements. Previous studies suggest that if the difference between the measured HR and the actual HR exceeds 10 units, the measurement is generally considered inaccurate~\cite{nelson2019accuracy}. We defined various thresholds $\tau = [10, 20, 25, 30, 31, 32, 33, 34, 35, 40]$ for the measurement error. System performance was evaluated using accuracy, calculated as $\text{Accuracy} = \left(\frac{TP}{TP + FP}\right) \times 100$, where $TP$ is the number of correctly identified unacceptable cases, and $FP$ is the number incorrectly classified as unacceptable.

\section{Experimental Results and Discussion}
\subsection{Results}
\label{sec:exp}

\begin{table}[!htbp]
\small
\setlength{\abovecaptionskip}{2pt}
\setlength{\belowcaptionskip}{2pt}
\centering
\begin{tabular}{l|ll}
\toprule
\textbf{$\tau$} & \textbf{Polar Accuracy} & \textbf{Garmin Accuracy} \\
\midrule
10 & 60.20  & 57.71  \\
20 & 85.90  & 84.21  \\
25 & 93.55  & 100.00 \\
30 & 91.67  & 100.00 \\
31 & 90.00  & 100.00 \\
32 & 83.33  & 100.00 \\
33 & 75.00  & 100.00 \\
34 & 100.00 & 100.00 \\
35 & 100.00 & 100.00 \\
40 & 100.00 & 100.00 \\
\bottomrule
\end{tabular}
\caption{Accuracy of the warning system on Polar and Garmin datasets across thresholds.}
\label{table:results}
\normalsize
\end{table}

We evaluated the performance of the warning system using data collected from the Polar and Garmin devices. Following the preprocessing steps and training procedures outlined earlier, the measurement error estimator was trained, and the performance of the warning system was assessed on both datasets. The results are summarized in Table~\ref{table:results}. Two thresholds, \( \tau_a = 20 \) and \( \tau_b = 34 \), were identified as particularly effective for both models. At the first threshold, \( \tau_a = 20 \), the model correctly detected approximately 86\% of unacceptable measurements in the Polar dataset. For the Garmin dataset, the model achieved a similar performance, correctly detecting about 84\% of unacceptable measurements at the same threshold value. When the second threshold, \( \tau_b = 34 \), was applied, the model detected 100\% of all unacceptable measurements in both datasets. To further evaluate the system performance, we analyzed its behavior across different error thresholds and tolerance levels on the Polar verity sense data, as shown in Table 2. The results demonstrate that at lower tolerance levels \(\tau = 3\), the system achieved F1 scores ranging from 85.5 to 86.6 with accuracy between 75.2\% and 76.2\%. As we increased the tolerance threshold to $\tau $= 5, both metrics improved, with F1 scores reaching 91.1 and accuracy increasing to 83.7\%. At the highest tolerance level tested \(\tau = 7\), the system showed its best performance with F1 scores of 93.5 and accuracy of 87.7\%.

\begin{table}[!htbp]
\small
\setlength{\abovecaptionskip}{2pt}
\setlength{\belowcaptionskip}{2pt}
\centering
\begin{tabular}{p{1.6cm}|p{0.7cm}p{0.7cm}|p{0.7cm}p{0.7cm}|p{0.7cm}p{0.7cm}}

\toprule
\textbf{GT vs Model Output Diff} & \multicolumn{2}{c|}{$\tau = 3$} & \multicolumn{2}{c|}{$\tau = 5$} & \multicolumn{2}{c}{$\tau = 7$} \\
\midrule
 & \textbf{F1} & \textbf{Acc} & \textbf{F1} & \textbf{Acc} & \textbf{F1} & \textbf{Acc} \\
\midrule
3  & 86.2 & 76.2 &      &      &      &      \\
4  & 86.1 & 75.8 &      &      &      &      \\
5  & 86.6 & 75.6 & 91.1 & 83.7 &      &      \\
6  & 85.9 & 75.5 & 91.1 & 83.6 &      &      \\
7  & 85.9 & 75.4 & 91.0 & 83.6 & 93.5 & 87.7 \\
8  & 85.9 & 75.4 & 91.0 & 83.5 & 93.5 & 87.7 \\
9  & 85.9 & 75.3 & 91.0 & 83.5 & 93.5 & 87.7 \\
10 & 85.9 & 75.3 & 90.9 & 83.5 & 93.5 & 87.7 \\
20 & 85.9 & 75.2 & 90.9 & 83.4 & 93.4 & 87.7 \\
25 & 85.9 & 75.2 & 90.9 & 83.4 & 93.4 & 87.7 \\
30 & 85.5 & 75.2 & 90.9 & 83.4 & 93.4 & 87.7 \\
\bottomrule
\end{tabular}
\caption{F1 score and accuracy for various thresholds on Polar dataset. "GT" = ground truth.}
\label{table:cutoff_results}
\normalsize
\end{table}

\subsection{Discussion}
Our formative study demonstrates promising initial outcomes in addressing the three key contributions (C1,C2,C3) outlined in the introduction. First, regarding the viability of using machine learning to predict periods of reduced HR accuracy, our early results support this approach. The model's performance at two critical thresholds ($\tau_a = 20$ and $\tau_b = 34$) suggests significant potential, with the system correctly identifying over 84\% of unacceptable measurements in the Garmin dataset and 86\% in Polar the dataset at $\tau_a = 20$. The architecture design prioritizes computational efficiency suitable for wearable devices while maintaining independence from original waveforms and specific hardware implementations. This approach theoretically enables broad applicability across different wearable platforms. The proposed three-tier traffic light system for communicating accuracy levels offers a potentially intuitive interface that could enhance user trust through transparency, although user studies are needed to validate this communication approach. The adaptability of the warning thresholds also suggests potential for customization across different use cases, though specific thresholds would need to be validated for each intended use case.

\subsection{Limitations}
While the system has demonstrated promising results, additional testing is needed to comprehensively assess its performance. This includes evaluating the system across a wider range of accuracy metrics and investigating its behavior under different conditions and thresholds. Another important consideration is the composition of the subject group. The study included 11 individuals who were taking heart rate-modifying medications. These medications could introduce variability in HR measurements that the model may find challenging to account for. The findings may also have limited generalizability to younger populations. The participants in the study were predominantly older adults, with a mean age of approximately 59 years. Future studies should include a broader age range to evaluate system effectiveness across different demographic groups. The model was developed and evaluated using data collected at a fixed sampling rate of 1 Hz. Wearable devices may operate at different sampling rates, and applying the system to such devices may require resampling strategies or adjustments to the model architecture. Another limitation is the system’s sensitivity to motion artifacts. Although the dataset includes real-world scenarios, the model does not explicitly account for noise introduced by body movement.

\subsection{Future Research Directions}
Future research could focus on refining the model architecture to enhance performance and reduce computational demands. Exploring the integration of additional physiological signals, such as accelerometer data, could also improve the system accuracy. Expanding the testing of the system across a broader range of devices and user populations would help to generalize the findings. Further research could explore diverse ways of communicating measurement accuracy to users. Effective communication of HR accuracy is crucial to ensure that users understand the reliability of the information users are receiving, potentially impacting their decision-making in health and fitness contexts. Colors may not be accessible to all users. Confidence scores, shown as percentages, can offer a more quantitative view of accuracy. While this approach is precise, users might not intuitively understand what different scores mean. Future studies could explore how users interpret confidence scores and how brief explanations or visuals might make these scores more meaningful. Contextual feedback—such as messages like ``Movement detected—HR accuracy may be reduced''—could further enhance usability by explaining why accuracy is compromised. This helps users understand accuracy drops and make adjustments, while it requires accurate error-source detection. Research into the effectiveness of these messages could identify which types of information users find most helpful and reassuring. Finally, displaying a graphical history of HR accuracy over time could help users recognize patterns and make adjustments to improve data quality. Though wearable screens are small, minimalist design could make trend displays feasible. 
\section{Conclusion}
This work presents the conceptual design and exploratory development and evaluation of a real-time warning system that aims to enhance user experience and trust in wearable health devices by detecting inaccuracies in PPG-based heart rate measurements. It addresses a key HCI challenge: bridging the gap between widespread adoption of wearables and the need for transparent, interpretable data / outcome measures. The system demonstrates that accuracy feedback can be delivered in real time without access to raw sensor data or proprietary algorithms. We hypothesize that implementing such a warning system can empower users with a more nuanced understanding of their physiological data, aligning with the goals of the HCI community of creating more transparent and trustworthy health technologies.

\label{sec:conc}
\bibliographystyle{ACM-Reference-Format}
\bibliography{ref}

\end{document}